\documentclass[aps,prb,preprint,groupedaddress]{revtex4-2}

\usepackage{amsmath}
\usepackage{graphicx}
\usepackage{epstopdf}
\DeclareGraphicsExtensions{.eps}

\begin{document}
\title{Exploring High-Temperature Superconductivity in the Extended Hubbard Model with Antiferromagnetic Tendencies}

\author{Zhipeng Sun}
\email{zpsun@csrc.ac.cn}
\affiliation{Beijing Computational Science Research Center, Beijing 100193, China}

\author{Hai-Qing Lin}
\email{haiqing0@csrc.ac.cn}
\affiliation{Beijing Computational Science Research Center, Beijing 100193, China}
\affiliation{Zhejiang University, Hangzhou 310027, China}

\date{\today}

\begin{abstract}
The enigma of unconventional superconductivity in doped cuprates presents a formidable challenge in the realm of condensed matter physics. Recent findings of strong near-neighbor attractions in one-dimensional cuprate chains suggest a new avenue for investigating cuprate superconductors. Consequently, we revisited the superconductivity in the extended Hubbard model at the mean-field level. Anticipating a prevalence of antiferromagnetic order due to strong local Coulomb repulsion, our calculations reveal the coexistence of superconducting and antiferromagnetic orders across a wide range of doping at sufficiently low temperatures. The mean-field results capture some key features of cuprate superconductors, including $d$-wave pairing symmetry, a dome-shaped dependence of $T_c$ on doping, and higher superconducting transition temperatures. Additionally, we observed a nearly proportional relationship between $T_c$ and the strength of the nearest-neighbor attraction, reminiscent of experimental findings at the FeSe/SrTiO3 interface. The mean-field results suggest that the extended Hubbard model could be the appropriate framework for investigating cuprate superconductivity and offer insights for more precise calculations within this model in future.

\end{abstract}

\maketitle

\section{Introduction}
The unconventional superconductivity in doped cuprate materials has been a focal point in condensed matter physics \cite{Micnas1990RMP,Tsuei2000RMP,Scalapino2012,Sobota2021RMP} since its discovery in 1986 \cite{Bednorz1986}. In contrast to conventional metal-based superconductors well-described by BCS theory \cite{Bardeen1957}, the unconventional nature of these materials manifests in various aspects such as narrow band electronic structures, superconducting (SC) transition temperatures ($T_c$) exceeding the McMillan limit \cite{McMillan1968}, and dominantly $d$-wave pairing symmetry \cite{Tsuei2000RMP}. The dome-shaped variation of $T_c$ with doping concentration, along with unusual isotope effects \cite{Zhao2001,Keller2005}, introduces additional complexities requiring explanation. Furthermore, above $T_c$, cuprate superconductors exhibit strongly correlated phenomena such as the pseudogap \cite{Timusk1999}, the stripe order \cite{Tranquada2020}, and the strange metallic behavior \cite{Phillips2022} that defy traditional Ginzburg-Landau framework, adding a layer of mystery to the mechanism.

Faced with these puzzles, the prevailing view among researchers is that superconductivity in cuprates does not arise from the electron-phonon mechanism advocated by BCS theory. Instead, there has been a shift towards exploring new pairing mechanisms. Based on the electronic structure of cuprates, their physics is believed to be describable by the Hubbard model or its extended versions \cite{Anderson1987,Scalapino2012}. A key and highly controversial question is whether superconductivity can exist in the simplest two-dimensional Hubbard model. On this matter, there are both negative and affirmative studies, yet a consensus remains elusive to date \cite{Lin1988,Bickers1989,Qin2020,Xu2023}. In the recent review \cite{Singh2021}, N. Singh summarized several leading theories supporting the existence of superconductivity in the Hubbard model. While these theories have achieved notable successes, there are also aspects that remain unsatisfactorily addressed. 

A crucial argument against the electron-phonon mechanism in cuprate superconductors is the McMillan limit \cite{Keimer2015}. However, it's important to clarify that the McMillan formula relies on two fundamental premises: first, the electronic density of states near the Fermi level is nearly constant, and second, the electron-phonon coupling constant is much smaller than the Debye frequency and the bandwidth. Regarding strong electron-phonon coupling, it is meaningful for superconductors under high pressure \cite{Gorkov2018}, but its relevance for exploring room pressure superconductivity may be limited. Regarding the density of states, in the presence of van Hove singularities (vHS), the dependence of $T_c$ on the electron-phonon coupling constant is much different from the BCS formula. The influence of vHS has been extensively discussed in numerous papers \cite{Abrikosov1993,Newns1995,Dagotto1995,Markiewicz2023}. 

The impact of divergent density of states on $T_c$ is elucidated more clearly in a short article \cite{Miyahara2007}. The authors explored a simple two-band BCS Hamiltonian, with one of the bands exhibiting a flat dispersion, within the mean-field approximation. They observed that $T_c$ is nearly proportional to the pairing coupling constant, in stark contrast to the conventional BCS exponential law. Building on this insight, some researchers have initiated investigations into the potential for room temperature superconductivity on flat-band systems \cite{Volovik2013,Heikkila2016}. 

The supportive evidence for electron-phonon mechanism appeared in the recent experiments on the one-dimensional cuprate chains \cite{Chen2021Science}. The experimental group reported the synthesis and spectroscopic analysis of the one-dimensional cuprate $\text{Ba}_{2-x}\text{Sr}_{x}\text{CuO}_{3+d}$ over a wide range of hole doping. The results of angle-resolved photoemission experiments fail to match predictions of the simple Hubbard model, while an additional strong near-neighbor attraction quantitatively explains experiments for all accessible doping levels. Such attraction may arise from the coupling to phonons, and there have been several works followed \cite{Wang2021PRL,Tang2023}. Nevertheless, as suggested by the group, the minimal model for cuprate superconductivity is likely the two-dimensional extended Hubbard model, which contains the nearest-neighbor attraction. Despite extensive research on superconductivity within this model before the experimental discovery \cite{Micnas1988PRB,Su2004,Tobijaszewska2005,Kheirkhah2020}, the attention has been reignited \cite{Jiang2022PRB,Peng2023}.

Given the lack of efficient and conclusive solvers for the general cases of the two-dimensional extended Hubbard model, we resorted to mean-field methods to revisit this model, acknowledging that the results may be subject to debate. Taking into account the strong local Coulomb repulsion, we assumed that the system has an antiferromagnetic (AFM) tendency. As a result, across a wide range of doping at sufficiently low temperatures, SC and AFM orders coexist. We found that the mean-field calculations can capture some features of cuprate superconductors, such as $d$-wave preference, a dome-shaped dependence of $T_c$ on doping, and higher $T_c$. Particularly striking was the discovery of an almost proportional relationship between $T_c$ and the strength of the nearest-neighbor attraction, which was evidenced at the FeSe/SrTiO3 interface \cite{Song2019}. We conjecture that this proportional relationship arises from the vHS near the Fermi energy. Our mean-field results suggest that the extended Hubbard model could be the appropriate framework for investigating cuprate superconductivity and are expected to offer insights for more precise calculations within this model in future.

This paper is structured as follows. In Section \ref{sec:method}, we outline the mean-field treatment of the extended Hubbard model and derive the gap equation near $T_c$. Next, in Section \ref{sec:numerical}, we compute $T_c$ over a broad range of parameters and examine its characteristics. Then in Section \ref{sec:discussions}, we compare our results with previous theoretical studies and experimental findings and engage in a discussion of the implications. Finally, a summary is made in Sec \ref{sec:summary}. 

\section{Model and Method}\label{sec:method}
We investigated the extended Hubbard model with strong local repulsion $U$ and nearest-neighbor attraction $\left|V\right|$ on a two-dimensional square $\mathcal{V}=L\times L$ lattice, whose Hamiltonian reads 
\begin{equation}
    \begin{aligned}
    \hat{H} &=-t\sum_{\alpha,\boldsymbol{r},\boldsymbol{\delta}}\hat{c}_{\alpha,\boldsymbol{r}+\boldsymbol{\delta}}^{\dagger}\hat{c}_{\alpha,\boldsymbol{r}}
    -\mu\sum_{\boldsymbol{r}} \hat{\rho}_{\boldsymbol{r}}
    \\ &\quad +U\sum_{\boldsymbol{r}}\hat{n}_{\uparrow,\boldsymbol{r}}\hat{n}_{\downarrow,\boldsymbol{r}}-\frac{\left|V\right|}{2}\sum_{\boldsymbol{r},\boldsymbol{\delta}}\hat{\rho}_{\boldsymbol{r}+\boldsymbol{\delta}}\hat{\rho}_{\boldsymbol{r}}.
    \end{aligned} \label{eq:model}
\end{equation}
Here $\hat{c}_{\alpha,\boldsymbol{r}}^{\dagger}$ ($\hat{c}_{\alpha,\boldsymbol{r}}$) is the fermionic creation (annihilation) operator with spin $\alpha$ at lattice site $\boldsymbol{r}$, $\hat{n}_{\alpha,\boldsymbol{r}}\equiv\hat{c}_{\alpha,\boldsymbol{r}}^{\dagger}\hat{c}_{\alpha,\boldsymbol{r}}$ is the spin-selective density operator, and $\hat{\rho}_{\boldsymbol{r}}=\hat{n}_{\uparrow,\boldsymbol{r}}+\hat{n}_{\downarrow,\boldsymbol{r}}$ is the charge density operator. $\boldsymbol{\delta}$ represents the vectors linking nearest neighbors, $t$ is the nearest-neighbor hopping strength, and $\mu$ is the chemical potential. 

\subsection{General mean-field framework}
Within the symmetry-broken Hartree-Fock framework, the local interacting term can be reduced to:
\begin{equation}
U\sum_{\boldsymbol{r}}\left(n_{\uparrow,\boldsymbol{r}}\hat{n}_{\downarrow,\boldsymbol{r}}+\hat{n}_{\uparrow,\boldsymbol{r}}n_{\downarrow,\boldsymbol{r}}\right)+U\sum_{\boldsymbol{r}}\left(\hat{c}{}_{\uparrow,\boldsymbol{r}}^{\dagger}\hat{c}{}_{\downarrow,\boldsymbol{r}}^{\dagger}\Delta_{\boldsymbol{r}}+\text{H.c.}\right).\label{eq:u-term-mf}
\end{equation}
Here, the local mean fields are defined as $n_{\alpha,\boldsymbol{r}}\equiv\left\langle \hat{n}_{\alpha,\boldsymbol{r}}\right\rangle $
and $\Delta_{\boldsymbol{r}}=\left\langle \hat{c}_{\downarrow,\boldsymbol{r}}\hat{c}_{\uparrow,\boldsymbol{r}}\right\rangle $.
In terms of the mean-field charge density $\rho_{\boldsymbol{r}}=n_{\uparrow,\boldsymbol{r}}+n_{\downarrow,\boldsymbol{r}}$
and the mean-field $z$-spin density $m_{\boldsymbol{r}}=\frac{1}{2}\left(n_{\uparrow,\boldsymbol{r}}-n_{\downarrow,\boldsymbol{r}}\right)$,
$n_{\alpha,\boldsymbol{r}}$ can be expressed as $\frac{1}{2}\rho_{\boldsymbol{r}}+\epsilon^{\alpha}m_{\boldsymbol{r}}$
with $\epsilon^{\uparrow}=1$ and $\epsilon^{\downarrow}=-1$.  ``H.c.'' represents for the Hermitian conjugation. The
mean-field approximation for the nonlocal interacting term is given by:
\begin{equation}
    \begin{aligned}
&-\left|V\right|\sum_{\boldsymbol{r},\boldsymbol{\delta}}\rho_{\boldsymbol{r}+\boldsymbol{\delta}}\hat{\rho}_{\boldsymbol{r}}+\left|V\right|\sum_{\alpha,\boldsymbol{r},\boldsymbol{\delta}}\hat{c}{}_{\alpha,\boldsymbol{r}+\boldsymbol{\delta}}^{\dagger}\hat{c}_{\alpha,\boldsymbol{r}}n_{\alpha,\boldsymbol{\delta},\boldsymbol{r}}^{\prime} \\ &\quad -\frac{\left|V\right|}{2}\sum_{\alpha,\alpha^{\prime},\boldsymbol{r},\boldsymbol{\delta}}\left(\hat{c}{}_{\alpha,\boldsymbol{r}+\boldsymbol{\delta}}^{\dagger}\hat{c}{}_{\alpha^{\prime},\boldsymbol{r}}^{\dagger}\Delta_{\alpha^{\prime},\alpha,\boldsymbol{\delta},\boldsymbol{r}}^{\prime}+\text{H.c.}\right).
    \end{aligned}
\label{eq:v-term-mf}
\end{equation}
Here, the nonlocal mean fields are defined as $n_{\alpha,\boldsymbol{\delta},\boldsymbol{r}}^{\prime}\equiv\left\langle \hat{c}_{\alpha,\boldsymbol{r}}^{\dagger}\hat{c}_{\alpha,\boldsymbol{r}+\boldsymbol{\delta}}\right\rangle $
and $\Delta_{\alpha^{\prime},\alpha,\boldsymbol{\delta},\boldsymbol{r}}^{\prime}\equiv\left\langle \hat{c}_{\alpha^{\prime},\boldsymbol{r}}\hat{c}_{\alpha,\boldsymbol{r}+\boldsymbol{\delta}}\right\rangle $.

Replacing the $U$ term by Eq. (\ref{eq:u-term-mf}) and $V$ term by Eq. (\ref{eq:v-term-mf}), we obtain the mean field Hamiltonian
$\hat{H}_{\text{MF}}=\hat{H}_{0}+\hat{H}_{\text{SC}}$. Here the ``normal'' part takes the form
\begin{equation}
\hat{H}_{0}=-\sum_{\alpha,\boldsymbol{r},\boldsymbol{\delta}}\tilde{t}_{\boldsymbol{r}, \boldsymbol{r}+\boldsymbol{\delta}}\hat{c}_{\alpha,\boldsymbol{r}+\boldsymbol{\delta}}^{\dagger}\hat{c}_{\alpha,\boldsymbol{r}}-\sum_{\alpha,\boldsymbol{r}}\tilde{\mu}_{\boldsymbol{r}}\hat{c}_{\alpha,\boldsymbol{r}}^{\dagger}\hat{c}_{\alpha,\boldsymbol{r}},\label{eq:ham0}
\end{equation}
with the effective hopping $\tilde{t}_{\boldsymbol{r}, \boldsymbol{r}+\boldsymbol{\delta}} = t - \left|V\right|n_{\alpha,\boldsymbol{\delta},\boldsymbol{r}}^{\prime}$, and effective chemical potential $\tilde{\mu}_{\boldsymbol{r}} = \mu + \sum_{\boldsymbol{\delta}}\left|V\right|\rho_{\boldsymbol{r}+\boldsymbol{\delta}} - \left(\frac{1}{2}\rho_{\boldsymbol{r}}-\epsilon^{\alpha}m_{\boldsymbol{r}}\right)U$. The SC term reads
\begin{equation}
\begin{aligned}
\hat{H}_{\text{SC}}&=\Big(U\sum_{\boldsymbol{r}}\hat{c}_{\uparrow,\boldsymbol{r}}^{\dagger}\hat{c}_{\downarrow,\boldsymbol{r}}^{\dagger}\Delta_{\boldsymbol{r}}-\left|V\right|\sum_{\boldsymbol{r},\boldsymbol{\delta}}\hat{c}_{\uparrow,\boldsymbol{r}+\boldsymbol{\delta}}^{\dagger}\hat{c}_{\downarrow,\boldsymbol{r}}^{\dagger}\Delta_{\boldsymbol{\delta},\boldsymbol{r}}^{\prime}\\&\quad -\frac{\left|V\right|}{2}\sum_{\alpha,\boldsymbol{r},\boldsymbol{\delta}}\hat{c}_{\alpha,\boldsymbol{r}+\boldsymbol{\delta}}^{\dagger}\hat{c}_{\alpha,\boldsymbol{r}}^{\dagger}\Delta_{\alpha,\boldsymbol{\delta},\boldsymbol{r}}^{t}\Big)+\text{H.c.}.
\end{aligned}
\label{eq:ham-sc}
\end{equation}
Here $\Delta_{\boldsymbol{\delta},\boldsymbol{r}}^{\prime}=\left\langle \hat{c}_{\downarrow,\boldsymbol{r}}\hat{c}_{\uparrow,\boldsymbol{r}+\boldsymbol{\delta}}\right\rangle $ and $\Delta_{\alpha,\boldsymbol{\delta},\boldsymbol{r}}^{t}=\left\langle \hat{c}_{\alpha,\boldsymbol{r}}\hat{c}_{\alpha,\boldsymbol{r}+\boldsymbol{\delta}}\right\rangle $ are the nonlocal unequal-spin and equal-spin pairing mean fields, respectively. 

As the mean field Hamiltonian $\hat{H}_{\text{MF}}$ is quadratic with respect to fermionic operators, thus enabling the establishment of the self-consistent equations for the mean fields. By solving the equations, the mean fields can be calculated, and desired physical quantities can be obtained. 

\subsection{AFM ansatz in the ``normal'' phase}
The mean field equations can yield various symmetry-broken solutions. However, due to the strong local repulsive interaction, we assume that above $T_c$, the system is either in the pure AFM phase or in the Fermi liquid phase. Under this assumption, the mean spin density takes the form $m_{\boldsymbol{r}}=m\text{e}^{\text{i}\boldsymbol{Q}\cdot\boldsymbol{r}}$ with  $\boldsymbol{Q}=\left(\pi,\pi\right)$ and $m\ge0$. In addition, we simplify the problem as much as possible by setting $\rho_{\boldsymbol{r}}=\rho$, and $n_{\alpha,\boldsymbol{\delta},\boldsymbol{r}}^{\prime}=\frac{1}{2}\rho^{\prime}$. The order parameters in the ``normal'' (non-SC) phase are then $\rho =\frac{1}{\mathcal{V}}\sum_{\alpha,\boldsymbol{k}}\left\langle \hat{c}_{\alpha,\boldsymbol{k}}^{\dagger}\hat{c}_{\alpha,\boldsymbol{k}}\right\rangle $, 
        $\rho^{\prime}  =\frac{1}{2\mathcal{V}}\sum_{\alpha,\boldsymbol{k}}\left\langle \hat{c}_{\alpha,\boldsymbol{k}}^{\dagger}\hat{c}_{\alpha,\boldsymbol{k}}\right\rangle \gamma_{\boldsymbol{k}}, $
        $m  =\frac{1}{2\mathcal{V}}\sum_{\alpha,\boldsymbol{k}}\epsilon^{\alpha}\left\langle \hat{c}_{\alpha,\boldsymbol{k}}^{\dagger}\hat{c}_{\alpha,\boldsymbol{k}+\boldsymbol{Q}}\right\rangle$.
Here $\gamma_{\boldsymbol{k}}=\cos k_{x}+\cos k_{y}$ is $s$-wave symmetric. 

The ``normal'' part Eq. (\ref{eq:ham0}) of the mean field Hamiltonian is then simplified as
\begin{equation}\hat{H}_{0}=\sum_{\alpha,\boldsymbol{k}}\left(-2\tilde{t}\gamma_{\boldsymbol{k}}-\tilde{\mu}\right)\hat{c}{}_{\alpha,\boldsymbol{k}}^{\dagger}\hat{c}_{\alpha,\boldsymbol{k}}-\sum_{\alpha,\boldsymbol{k}}\epsilon^{\alpha}Um\hat{c}{}_{\alpha,\boldsymbol{k}}^{\dagger}\hat{c}_{\alpha,\boldsymbol{k}+\boldsymbol{Q}},\label{eq:ham0-r}
\end{equation}
where $\tilde{t}=t-\frac{1}{2}\left|V\right|\rho^{\prime}$ is the renormalized nearest-neighbor hopping strength and $\tilde{\mu}=\mu-\frac{1}{2}U\rho+4\left|V\right|\rho$ is the renormalized chemical potential. In absence of SC orders where $\hat{H}_{\text{MF}}=\hat{H}_{0}$, we can introduce the Bogoliubov transformation:
\begin{subequations}\label{eq:trans}
\begin{equation}
    \left[\begin{array}{c}
    \hat{c}_{\uparrow,-\boldsymbol{p}}\\
    \hat{c}_{\uparrow,-\boldsymbol{p}+\boldsymbol{Q}}
    \end{array}
    \right]
    =\left[\begin{array}{cc}
    \cos\frac{\theta_{\boldsymbol{p}}}{2} & -\sin\frac{\theta_{\boldsymbol{p}}}{2}\\
    \sin\frac{\theta_{\boldsymbol{p}}}{2} & \cos\frac{\theta_{\boldsymbol{p}}}{2}\\
    \end{array}\right]\left[\begin{array}{c}
    \hat{a}_{1,\boldsymbol{p}}\\
    \hat{a}_{2,\boldsymbol{p}}
    \end{array}\right],
\end{equation}
\begin{equation}
    \left[\begin{array}{c}
    \hat{c}_{\downarrow,\boldsymbol{p}}^{\dagger}\\
\hat{c}_{\downarrow,\boldsymbol{p}+\boldsymbol{Q}}^{\dagger}
    \end{array}\right]=\left[\begin{array}{cc}
    \cos\frac{\theta_{\boldsymbol{p}}}{2} & \sin\frac{\theta_{\boldsymbol{p}}}{2}\\
    -\sin\frac{\theta_{\boldsymbol{p}}}{2} & \cos\frac{\theta_{\boldsymbol{p}}}{2}
    \end{array}\right]\left[\begin{array}{c}
    \hat{a}_{3,\boldsymbol{p}}\\
    \hat{a}_{4,\boldsymbol{p}}
    \end{array}\right].
\end{equation}
\end{subequations}
Here $\boldsymbol{p}$ is confined in the half of the first Brillouin zone, and $\theta_{\boldsymbol{p}}\in\left[0,\frac{\pi}{2}\right)$ is $s$-wave symmetric, determined by $\cot\theta_{\boldsymbol{p}}=\frac{2\tilde{t}}{Um}\gamma_{\boldsymbol{p}}$ if $m\not = 0$ otherwise $\theta_{\boldsymbol{p}}\equiv0$. By virtue of Eq. (\ref{eq:trans}), $\hat{H}_{0}$ is diagonalized as $\hat{H}_{0} = \sum_{i,\boldsymbol{p}} \tilde{h}^{i}_{\boldsymbol{p}} \hat{a}_{i,\boldsymbol{p}}^{\dagger} \hat{a}_{i,\boldsymbol{p}} $, where $\tilde{h}^{i}_{\boldsymbol{p}} = \left[\xi_{\boldsymbol{p},-}, \xi_{\boldsymbol{p},+}, -\xi_{\boldsymbol{p},-}, -\xi_{\boldsymbol{p},+}\right]$, with the effective disperions $\xi_{\boldsymbol{p},\pm}=\pm\sqrt{4\tilde{t}^{2}\gamma_{\boldsymbol{p}}^{2}+U^{2}m^{2}}-\tilde{\mu}$.
The ``normal'' mean fields above the critical temperature $T_{c}$ can then be solved from mean-field equations. 

\subsection{Ansatz for SC order parameters}
Taking into account the existence of AFM order, we make the following ansatz for SC order parameters: the local $\Delta_{\boldsymbol{r}}=\Delta$, the nonlocal equal-spin $\Delta_{\alpha,\boldsymbol{\delta},\boldsymbol{r}}^{t}=0$, and the nonlocal unequal-spin $\Delta_{\boldsymbol{\delta},\boldsymbol{r}}^{\prime}=\Delta_{\boldsymbol{\delta},\boldsymbol{0}}^{\prime}+\Delta_{\boldsymbol{\delta},\boldsymbol{Q}}^{\prime}\text{e}^{\text{i}\boldsymbol{Q}\cdot\boldsymbol{r}}$. The order parameters involved are given by
$\Delta =\frac{1}{\mathcal{V}}\sum_{\boldsymbol{k}}\left\langle \hat{c}_{\downarrow,\boldsymbol{k}}\hat{c}_{\uparrow,-\boldsymbol{k}}\right\rangle$ , $\Delta_{\boldsymbol{\delta},\boldsymbol{0}}^{\prime}  =\frac{1}{\mathcal{V}}\sum_{\boldsymbol{k}}\left\langle \hat{c}_{\downarrow,\boldsymbol{k}}\hat{c}_{\uparrow,-\boldsymbol{k}}\right\rangle \text{e}^{\text{i}\boldsymbol{k}\cdot\boldsymbol{\delta}}$,    $\Delta_{\boldsymbol{\delta},\boldsymbol{Q}}^{\prime}  =\frac{1}{\mathcal{V}}\sum_{\boldsymbol{k}}\left\langle \hat{c}_{\downarrow,\boldsymbol{k}+\boldsymbol{Q}}\hat{c}_{\uparrow,-\boldsymbol{k}}\right\rangle \text{e}^{\text{i}\boldsymbol{k}\cdot\boldsymbol{\delta}}$. The SC part Eq. (\ref{eq:ham-sc}) of the mean field Hamiltonian is then simplified as: 
\begin{equation}
\begin{aligned}
\hat{H}_{\text{SC}}&=\Big(U\sum_{\boldsymbol{k}}\hat{c}{}_{\uparrow,-\boldsymbol{k}}^{\dagger}\hat{c}{}_{\downarrow,\boldsymbol{k}}^{\dagger}\Delta-\left|V\right|\sum_{\boldsymbol{\delta},\boldsymbol{k}}\hat{c}{}_{\uparrow,-\boldsymbol{k}}^{\dagger}\hat{c}{}_{\downarrow,\boldsymbol{k}}^{\dagger}\Delta_{\boldsymbol{\delta},\boldsymbol{0}}^{\prime}\text{e}^{-\text{i}\boldsymbol{k}\cdot\boldsymbol{\delta}}\\&\quad -\left|V\right|\sum_{\boldsymbol{k}}\hat{c}{}_{\uparrow,-\boldsymbol{k}}^{\dagger}\hat{c}{}_{\downarrow,\boldsymbol{k}+\boldsymbol{Q}}^{\dagger}\Delta_{\boldsymbol{\delta},\boldsymbol{Q}}^{\prime}\text{e}^{-\text{i}\boldsymbol{k}\cdot\boldsymbol{\delta}}\Big)+\text{H.c.}.
\end{aligned}
    \label{eq:ham-sc-r}
\end{equation}

With Eqs. (\ref{eq:ham0-r}, \ref{eq:ham-sc-r}), the mean-field Hamiltonian takes the quadratic form $\hat{H}_{\text{MF}}=\sum_{i,j,\boldsymbol{p}}\hat{c}_{\boldsymbol{p}}^{i\dagger}h{}_{\boldsymbol{p}}^{i,j}\hat{c}_{\boldsymbol{p}}^{j}$,
where $\hat{c}{}_{\boldsymbol{p}}^{j\dagger}=\left[\begin{array}{cccc} \hat{c}{}_{\uparrow,-\boldsymbol{p}}^{\dagger} & \hat{c}_{\downarrow,-\boldsymbol{p}} & \hat{c}{}_{\uparrow,\boldsymbol{p}+\boldsymbol{Q}}^{\dagger} & \hat{c}_{\downarrow,\boldsymbol{p}+\boldsymbol{Q}} \end{array}\right]$ and $h_{\boldsymbol{p}}^{i,j}$ is a $4\times4$ matrix shown as
below:
\begin{widetext}
\begin{equation}
h_{\boldsymbol{p}}^{i,j}=\left[\begin{array}{cccc}
-2\tilde{t}\gamma_{\boldsymbol{p}}-\tilde{\mu} & -Um & U\Delta-\left|V\right|X_{\boldsymbol{0},\boldsymbol{p}} & -\left|V\right|X_{\boldsymbol{Q},\boldsymbol{p}}\\
-Um & 2\tilde{t}\gamma_{\boldsymbol{p}}-\tilde{\mu} & \left|V\right|X_{\boldsymbol{Q},\boldsymbol{p}} & U\Delta+\left|V\right|X_{\boldsymbol{0},\boldsymbol{p}}\\
U\Delta^{\ast}-\left|V\right|X_{\boldsymbol{0},\boldsymbol{p}}^{\ast} & \left|V\right|X_{\boldsymbol{Q},\boldsymbol{p}}^{\ast} & 2\tilde{t}\gamma_{\boldsymbol{p}}+\tilde{\mu} & -Um\\-\left|V\right| X_{\boldsymbol{Q},\boldsymbol{p}}^{\ast} & U\Delta^{\ast}+\left|V\right| X_{\boldsymbol{0},\boldsymbol{p}}^{\ast} & -Um & -2\tilde{t}\gamma_{\boldsymbol{p}}+\tilde{\mu}
\end{array}\right].\label{eq:h-mat}
\end{equation}
\end{widetext}
Here the order parameters $X_{\boldsymbol{0},\boldsymbol{p}}$ and $X_{\boldsymbol{Q},\boldsymbol{p}}$
are defined as 
\begin{equation}
X_{\boldsymbol{0},\boldsymbol{p}}  =\sum_{\boldsymbol{\delta}}\Delta_{\boldsymbol{\delta},\boldsymbol{0}}^{\prime}\text{e}^{-\text{i}\boldsymbol{p}\cdot\boldsymbol{\delta}}, \quad
X_{\boldsymbol{Q},\boldsymbol{p}} =\sum_{\boldsymbol{\delta}}\Delta_{\boldsymbol{\delta},\boldsymbol{Q}}^{\prime}\text{e}^{-\text{i}\boldsymbol{p}\cdot\boldsymbol{\delta}}.\label{eq:x-def}
\end{equation}
Note that both $X_{\boldsymbol{0},\boldsymbol{p}}$ and $X_{\boldsymbol{Q},\boldsymbol{p}}$ have only \textit{four} independent components.

With the Bogoliubov transformation Eq. (\ref{eq:trans}), the mean field Hamiltonian can be expressed as $\hat{H}_{\text{MF}}=\sum_{i,j,\boldsymbol{p}}\hat{a}_{i,\boldsymbol{p}}^{\dagger}\tilde{h}_{\boldsymbol{p}}^{i,j}\hat{a}_{j,\boldsymbol{p}}$ with $\tilde{h}_{\boldsymbol{p}}^{i,j}$ given by the following matrix: 
\begin{equation}
\tilde{h}_{\boldsymbol{p}}^{i,j}=\left[\begin{array}{cccc}
\xi_{\boldsymbol{p},-} & 0 & A_{\boldsymbol{p}} & B_{\boldsymbol{p}}\\
0 & \xi_{\boldsymbol{p},+} & -B_{\boldsymbol{p}} & C_{\boldsymbol{p}}\\
A_{\boldsymbol{p}}^{\ast} & -B_{\boldsymbol{p}}^{\ast} & -\xi_{\boldsymbol{p},-} & 0\\
B_{\boldsymbol{p}}^{\ast} & C_{\boldsymbol{p}}^{\ast} & 0 & -\xi_{\boldsymbol{p},+}
\end{array}\right]. \label{eq:ham-in-a}
\end{equation}
The undetermined variables $A_{\boldsymbol{p}}$, $B_{\boldsymbol{p}}$, $C_{\boldsymbol{p}}$ are given by
\begin{subequations}\label{eq:abc}
\begin{align}
    A_{\boldsymbol{p}} & =-\left|V\right|X_{\boldsymbol{0},\boldsymbol{p}}+\left|V\right|\sin\theta_{\boldsymbol{p}}X_{\boldsymbol{Q},\boldsymbol{p}}+U\cos\theta_{\boldsymbol{p}}\Delta, \\
    B_{\boldsymbol{p}} & =U\sin\theta_{\boldsymbol{p}}\Delta-\left|V\right|\cos\theta_{\boldsymbol{p}}X_{\boldsymbol{Q},\boldsymbol{p}}, \\
    C_{\boldsymbol{p}} & =\left|V\right|X_{\boldsymbol{0},\boldsymbol{p}}+\left|V\right|\sin\theta_{\boldsymbol{p}}X_{\boldsymbol{Q},\boldsymbol{p}}+U\cos\theta_{\boldsymbol{p}}\Delta.
    \end{align}
\end{subequations}
After diagonalizing the matrix $\tilde{h}_{\boldsymbol{p}}$, we can construct the SC gap equations.

\subsection{SC gap equations at the critical point}
Near the critical temperature $T_c$, the parameters $A_{\boldsymbol{p}}$, $B_{\boldsymbol{p}}$, and $C_{\boldsymbol{p}}$ are all small, so all quantities can be approximated to first order with respect to them. The SC gap equations can be derived and are shown as follows:
\begin{subequations}\label{eq:gap-eq}
\begin{equation}
\begin{aligned}
    \Delta & =-\frac{1}{\mathcal{V}}\sum_{\boldsymbol{p}}\Big(\left(A_{\boldsymbol{p}}F_{\boldsymbol{p}}^{-}+C_{\boldsymbol{p}}F_{\boldsymbol{p}}^{+}\right)\cos\theta_{\boldsymbol{p}}\\&\quad +2B_{\boldsymbol{p}}F_{\boldsymbol{p}}^{\prime}\sin\theta_{\boldsymbol{p}}\Big),
\end{aligned}
\end{equation}
\begin{equation}
\Delta_{\boldsymbol{\delta},\boldsymbol{0}}^{\prime}  =-\frac{1}{\mathcal{V}}\sum_{\boldsymbol{p}}\text{e}^{\text{i}\boldsymbol{p}\cdot\boldsymbol{\delta}}\left(A_{\boldsymbol{p}}F_{\boldsymbol{p}}^{-}-C_{\boldsymbol{p}}F_{\boldsymbol{p}}^{+}\right),
\end{equation}
\begin{equation}
    \begin{aligned}        \Delta_{\boldsymbol{\delta},\boldsymbol{Q}}^{\prime} & =\frac{1}{\mathcal{V}}\sum_{\boldsymbol{p}}\text{e}^{\text{i}\boldsymbol{p}\cdot\boldsymbol{\delta}}\Big(\left(A_{\boldsymbol{p}}F_{\boldsymbol{p}}^{-}+C_{\boldsymbol{p}}F_{\boldsymbol{p}}^{+}\right)\sin\theta_{\boldsymbol{p}}\\&\quad -2B_{\boldsymbol{p}}F_{\boldsymbol{p}}^{\prime}\cos\theta_{\boldsymbol{p}}\Big).
\end{aligned}
\end{equation}
\end{subequations}
Here the bubble diagram $F$'s are defined as:
\begin{align}
    F_{\boldsymbol{p}}^{\pm} & =\frac{1}{2\xi_{\boldsymbol{p},\pm}}\tanh\frac{\beta\xi_{\boldsymbol{p},\pm}}{2},\nonumber \\
    F_{\boldsymbol{p}}^{\prime} & =\frac{1}{\xi_{\boldsymbol{p},+}+\xi_{\boldsymbol{p},-}}\frac{1}{2}\left(\tanh\frac{\beta\xi_{\boldsymbol{p},-}}{2}+\tanh\frac{\beta\xi_{\boldsymbol{p},+}}{2}\right).\label{eq:F}
\end{align}

By combining Eqs. (\ref{eq:x-def}, \ref{eq:abc}, \ref{eq:gap-eq}), we obtain a \textit{ninth}-order homogeneous linear equation set. Based on the condition for the existence of a non-zero solution to the equation set, we can determine $T_c$ and hence explore the physical properties at the SC boundary.

\subsection{Reduction of gap equations by symmetries}
As mentioned above, both $X_{\boldsymbol{0},\boldsymbol{p}}$ and $X_{\boldsymbol{Q},\boldsymbol{p}}$ defined by Eq. (\ref{eq:x-def}) have only \textit{four} independent components; they can be decomposed as:
\begin{align}
    X_{\boldsymbol{0},\boldsymbol{p}} & =X_{\boldsymbol{0},\gamma}\gamma_{\boldsymbol{p}}+X_{\boldsymbol{0},\eta}\eta_{\boldsymbol{p}}+X_{\boldsymbol{0},\nu^{+}}\nu_{\boldsymbol{p}}^{+}+X_{\boldsymbol{0},\nu^{-}}\nu_{\boldsymbol{p}}^{-},\nonumber \\
    X_{\boldsymbol{Q},\boldsymbol{p}} & =X_{\boldsymbol{Q},\gamma}\gamma_{\boldsymbol{p}}+X_{\boldsymbol{Q},\eta}\eta_{\boldsymbol{p}}+X_{\boldsymbol{Q},\nu^{+}}\nu_{\boldsymbol{p}}^{+}+X_{\boldsymbol{Q},\nu^{-}}\nu_{\boldsymbol{p}}^{-}.\label{eq:x-ansatz}
\end{align}
Here, $\gamma_{\boldsymbol{p}}=\cos p_{x}+\cos p_{y}$ exhibits $s$-wave symmetry, $\eta_{\boldsymbol{p}}=\cos p_{x}-\cos p_{y}$ exhibits $d$-wave symmetry, and $\nu_{\boldsymbol{p}}^{\pm}=\sin p_{x}\pm\sin p_{y}$ exhibits $p$-wave symmetry. Due to these symmetries, the system of equations for $\Delta$, $X_{\boldsymbol{0},\varphi}$, and $X_{\boldsymbol{Q},\varphi}$ ($\varphi=\gamma,\eta,\nu^{+},\nu^{-}$) can be decomposed into \textit{four} smaller subsystems.

For pure $d$-wave or $p$-wave, the system of equations takes the form
\begin{equation}
    \left[\begin{array}{c}
    X_{\boldsymbol{0},\varphi}\\
    X_{\boldsymbol{Q},\varphi}
    \end{array}\right]=\left[\begin{array}{cc}
    M^{11} & M^{12}\\
    M^{21} & M^{22}
    \end{array}\right]\left[\begin{array}{c}
    X_{\boldsymbol{0},\varphi}\\
    X_{\boldsymbol{Q},\varphi}
    \end{array}\right].\label{eq:d-p-waves}
\end{equation}
Here the elements of the $2\times 2$ matrix $M$ are given by
\begin{equation}
    \begin{aligned}
    M^{11} &= \frac{\left|V\right|}{\mathcal{V}}\sum_{\boldsymbol{p}}\varphi_{\boldsymbol{p}}^{2} \left(F_{\boldsymbol{p}}^{-}+F_{\boldsymbol{p}}^{+}\right), \\
    M^{12} &= M^{21} = - \frac{\left|V\right|}{\mathcal{V}}\sum_{\boldsymbol{p}}\varphi_{\boldsymbol{p}}^{2} \left(F_{\boldsymbol{p}}^{-}-F_{\boldsymbol{p}}^{+}\right)\sin\theta_{\boldsymbol{p}}, \\
    M^{22} &= \frac{\left|V\right|}{\mathcal{V}}\sum_{\boldsymbol{p}}\varphi_{\boldsymbol{p}}^{2} \left(\sin^{2}\theta_{\boldsymbol{p}}\left(F_{\boldsymbol{p}}^{-}+F_{\boldsymbol{p}}^{+}\right)+2\cos^{2}\theta_{\boldsymbol{p}}F_{\boldsymbol{p}}^{\prime}\right).
    \end{aligned}
\end{equation}
Based on the condition for the existence of a non-zero solution to Eq. (\ref{eq:d-p-waves}), we can obtain $T_{c}$ for the onset of $d$-wave and $p$-wave pairing instabilities. Additionally, it is worth noting that $X_{\boldsymbol{Q},\varphi} \neq 0$ implies that the nonlocal SC order is subject to spatial modulation. The gap equation for pure $s$-wave pairing is put in Appendix. 

\subsection{Summary of mean-field formulation}
Through a general mean-field treatment of the extended Hubbard model's Hamiltonian Eq. (\ref{eq:model}), we obtain the mean-field Hamiltonian $\hat{H}_{\text{MF}} = \hat{H}_0 + \hat{H}_{\text{SC}}$, where $\hat{H}_0$ is given by Eq. (\ref{eq:ham0}) and $\hat{H}_{\text{SC}}$ is given by Eq. (\ref{eq:ham-sc}). Further assumptions lead to the simplification of $\hat{H}_{\text{MF}}$ into the form $\hat{H}_{\text{MF}} = \hat{c}^{i,\dagger}_{\boldsymbol{p}} h^{i,j}_{\boldsymbol{p}} c^{j}_{\boldsymbol{p}}$, where $h$ is a $4\times 4$ matrix as given by Eq. (\ref{eq:h-mat}). Near the critical temperature $T_c$, we derive the SC gap equation, provided by Eq. (\ref{eq:gap-eq}). Finally, considering the spatial symmetry of the SC order parameter, the SC gap equation is simplified to Eq. (\ref{eq:d-p-waves}) for both $d$-wave and $p$-wave cases.

\section{Numerical Results}\label{sec:numerical}
\subsection{Numerical implementation}
The numerical calculations are all performed on a $512\times 512$ square lattice. We set $t=1$ as the unit of energy, and only consider the case $\rho < 1$. The parameter range in consideration is based on the experimental results on the cuprate chains \cite{Chen2021Science}, where $U \sim 8t$ and $|\mathrm{V}| \sim t$. Considering that there is not enough experimental evidence to indicate the presence of strong near-neighbor attractions in the two-dimensional cuprates, we will consider a wider range of values for $\left|V\right|$.

In details of implementation, we treat $\mu^{\prime}=\tilde{\mu}+Um$ as a tunable parameter. For given $\mu^{\prime}$, we employ the bisection method to find the temperature $T_c$ at which the largest eigenvalue of the 2x2 matrix $M$, corresponding to Eq. (\ref{eq:d-p-waves}), equals $1$. Then other quantities of interest can be obtained. 

\subsection{Factors influencing $T_c$}

\begin{figure}[htbp]
    \centering
    \includegraphics[width=0.9\linewidth]{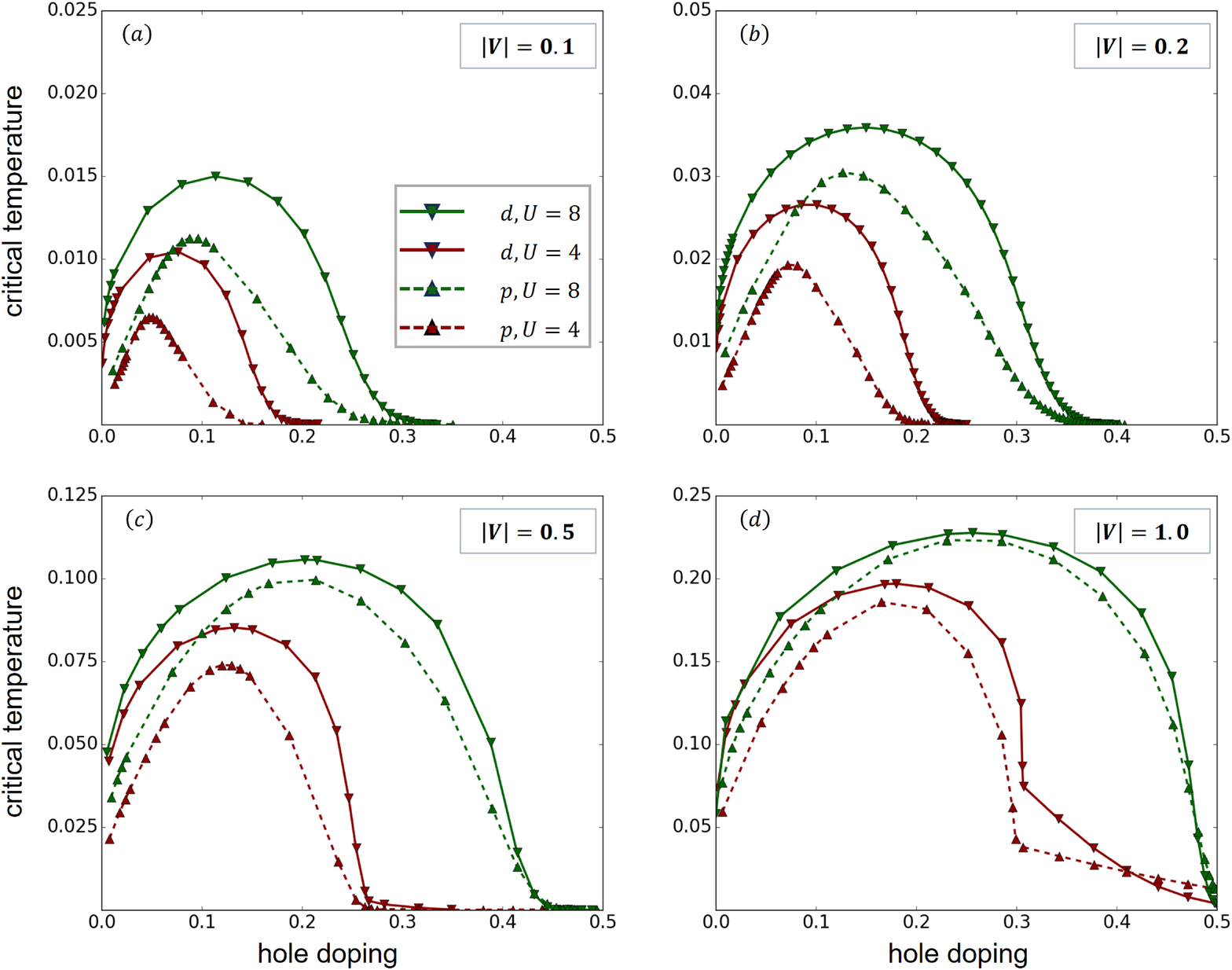}
    \caption{Critical temperature versus hole doping for $d$-wave instability and $p$-wave instability at varying values of $U$ (4.0, 8.0) and $\left\vert V\right\vert$ (0.1, 0.2, 0.5, 1.0). }
    \label{fig:tc_versus_doping}
\end{figure}

We focus on factors influencing $T_c$. Firstly, we plot $T_c$ for $d$-wave and $p$-wave pairing instabilities versus the doping $1-\rho$ in Fig. \ref{fig:tc_versus_doping}. Here we choose two set of $U$ values ($4.0$, $8.0$) and four sets of $|V|$ values ($0.1$, $0.2$, $0.5$, $1.0$). The plot exhibits \textit{three} key features of $T_c$:  the curve of $T_c$ versus doping forms a dome shape; $T_c$ is positively correlated with local repulsion $U$, but the relationship is not strong; $T_c$ demonstrates a positive correlation with the nearest-neighbor attraction $\left|V\right|$, and the dependence is statistically significant. Besides, $d$-wave instability is generally more prevalent than $p$-wave instability. 

\begin{figure}[htbp]
    \centering
    \includegraphics[width=0.9\linewidth]{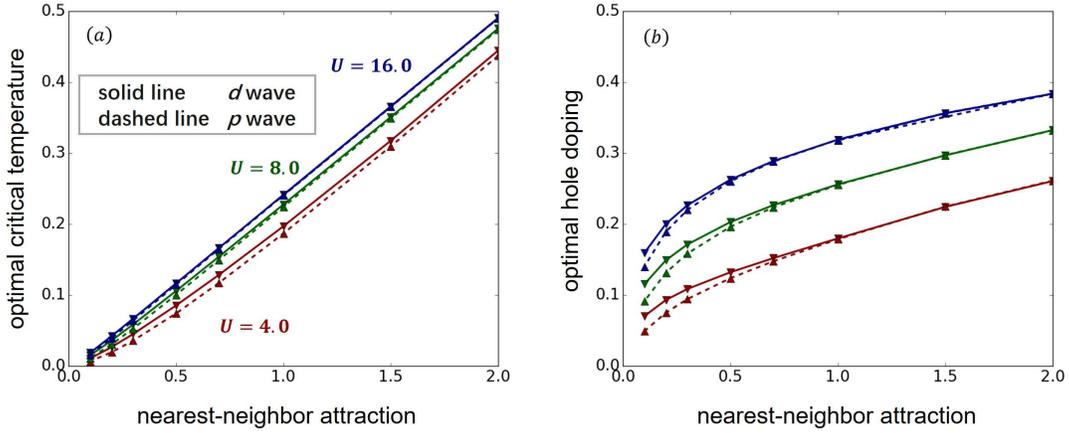}
    \caption{Optimal critical temperature and hole doping versus nearest-neighbor attraction for $d$-wave instability and $p$-wave instability at varying values of $U$ (4.0, 8.0, 16.0). }
    \label{fig:tc_versus_absv}
\end{figure}

To further investigate the dependence of $T_c$ on the nearest-neighbor attraction $\left|V\right|$, we plot the optimal $T_{c}$ versus $\left|V\right|$  at varying values of $U$ (4.0, 8.0, 16.0) in Fig. \ref{fig:tc_versus_absv}, together with the optimal doping. Notably, $T_{c}$ is almost proportional to $\left|V\right|$, and $T_{c} \sim 0.25\left|V\right|$ when $U$ and $\left|V\right|$ are both sufficiently large. This linear relationship is totally different from the exponential relationship in conventional superconductors. 

\begin{figure}[htbp]
    \centering
    \includegraphics[width=0.9\linewidth]{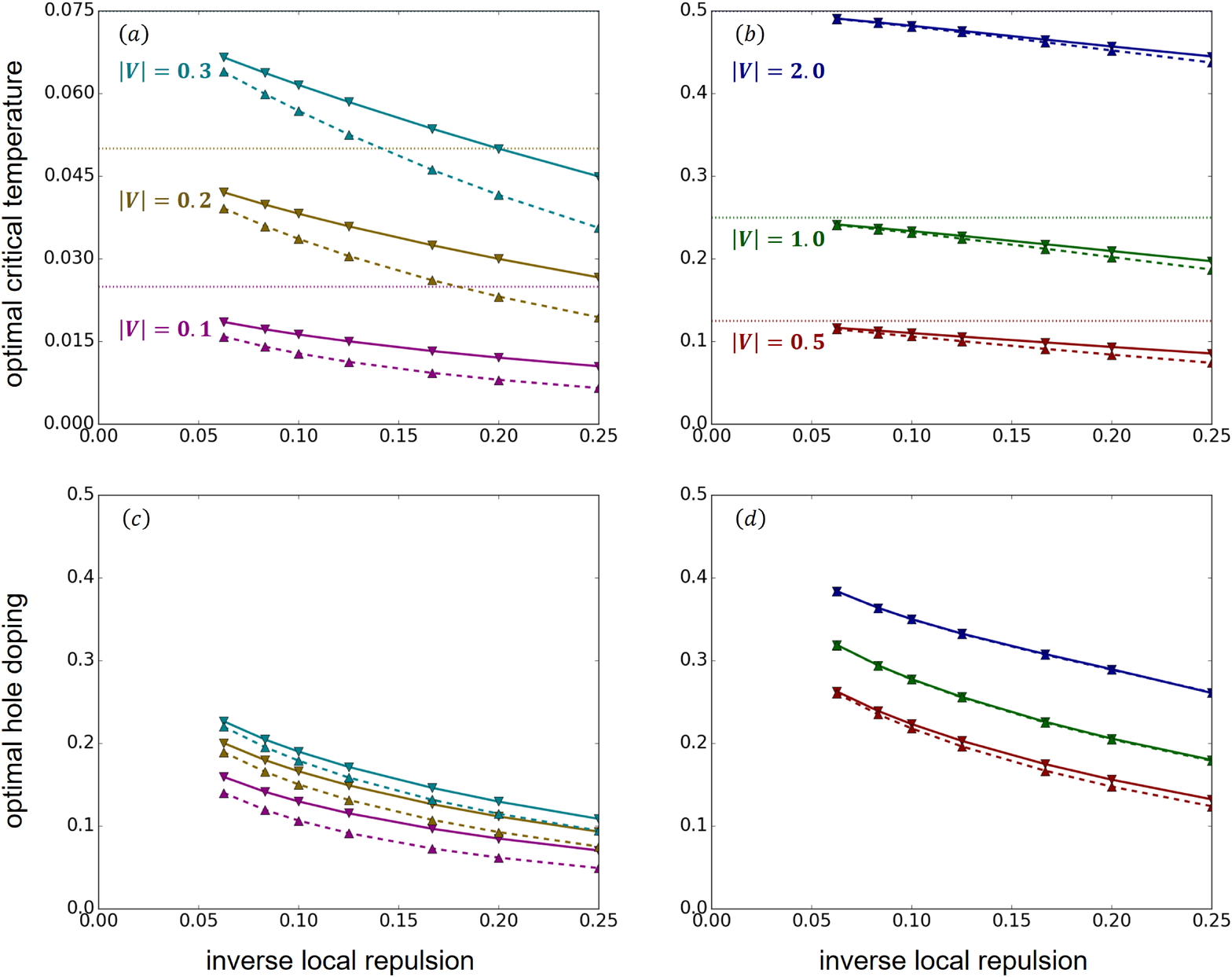}
    \caption{Optimal critical temperature and hole doping versus the inverse local repulsion for $d$-wave instability (solid line) and $p$-wave instability (dashed line) at varying values of $\left|V\right|$ (0.1, 0.2, 0.3, 0.5, 1.0, 2.0). }
    \label{fig:tc_versus_u}
\end{figure}

We also plot the optimal $T_c$ versus the inverse local repulsion at varying values of $\left|V\right|$ ($0.1$, $0.2$, $0.3$, $0.5$, $1.0$, $2.0$) in Fig. \ref{fig:tc_versus_u}, together with the optimal doping. It can be observed that increasing the local repulsion can enhance $T_{c}$. This enhancement is more significant when $\left|V\right|$ is relatively small, while it becomes less pronounced for strong nearest-neighbor attraction. 

\subsection{Possible origin of high $T_c$: role of density of states}
\begin{figure}[htbp]
    \centering
    \includegraphics[width=0.9\linewidth]{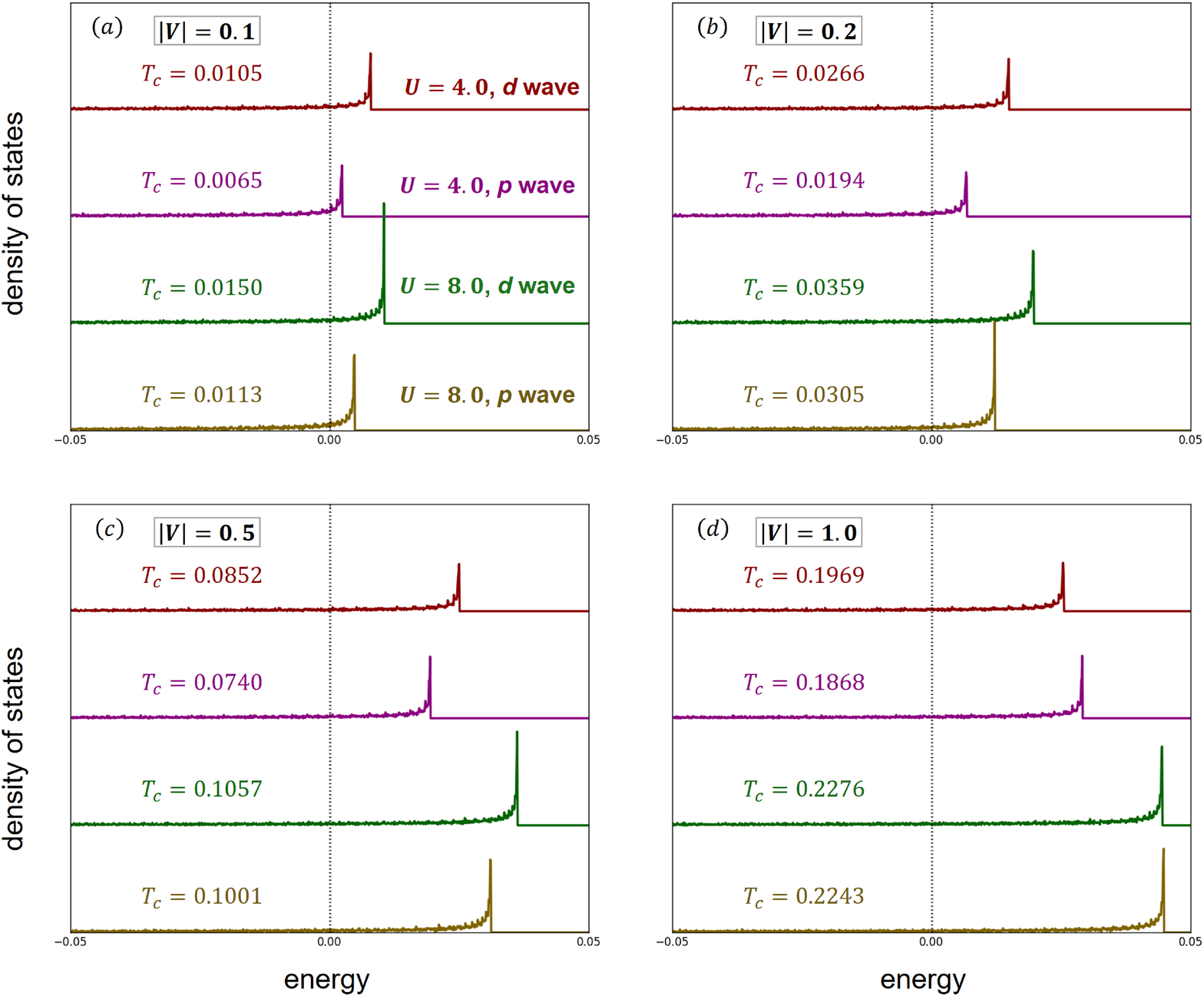}
    \caption{The density of states for several sets of model parameters. }
    \label{fig:dos}
\end{figure}

It is important to emphasize that the parameter $U$ does not manifest explicitly in the gap equation. Instead, its influence on the critical temperature $T_c$ is mediated by its effects on the electronic band structure. To elucidate this mechanism, we present the density of states for various parameter sets (at optimal dopings), depicted in Fig. \ref{fig:dos}. Notably, the density of states reveals a pronounced vHS situated in close proximity to the Fermi energy. This proximity potentially underpins the heightened $T_c$ observed in the extended Hubbard model. 

To understand this relationship, we simplify the gap equation Eq. (\ref{eq:d-p-waves}) in the limiting AFM case where $Um\gg 4\tilde{t}$. The gap equation at $T_c$ is then reduced to 
\begin{equation}
    \frac{1}{\left|V\right|} = \frac{1}{\mathcal{V}/2} \sum_{\boldsymbol{p}} \varphi_{\boldsymbol{p}}^{2} F_{\boldsymbol{p}}^{-}.
\end{equation}
Due to the presence of vHS, we derive an estimation for $T_c$,
\begin{equation}
    T_{c} \sim \frac{N_{0}\left|V\right|}{4} \frac{x}{\textrm{arctanh }x}, \label{eq:estimate}
\end{equation}
with $x=2\omega_{0}/N_{0}\left|V\right|$. Here $N_{0}$ represents the ratio of the number of states within the energy range $\left(-\omega_{0}, \omega_{0}\right)$ to that in the lower energy band, and $\omega_{0}$ can be set as the energy difference between the vHS and the Fermi energy. Note that Eq. (\ref{eq:estimate}) is applicable only when $N_{0} \left|V\right|>2\omega_{0}$. In the case $N_{0} \left|V\right| \gg 2\omega_{0}$, $T_{c} \sim N_{0}\left|V\right|/4$; that is, $T_{c}$ is proportional to $\left|V\right|$. 

\section{Discussions}\label{sec:discussions}
It is essential to acknowledge that mean-field methods have significant limitations when applied to two-dimensional Hubbard-like models. To address this, we reevaluate our numerical findings in light of existing experimental and theoretical results. 

\textit{Dome-shaped feature}. The critical temperature of hole-doped cuprate superconductors indeed exhibits an dome-shaped dependence on doping \cite{Sobota2021RMP}. However, in the case of lightly doping, cuprates do not exhibit SC behavior, which contrasts with our numerical results. In comparison to Micnas' work \cite{Micnas1988PRB}, where the neglect of $U$ resulted in the highest $T_c$ occurring at half-filling, we infer that the emergence of the dome-shaped feature is due to the influence of $U$. In our study, the effect of $U$ is manifested in the density of states; however, at higher orders of perturbation theory, $U$ would introduce corrections to the vertices. The absence of corrections to $U$ might account for the deviations between our numerical results and existing experimental findings.

\textit{$d$-wave preference}. Many hole-doped cuprate superconductors exhibit $d$-wave features \cite{Tsuei2000RMP}, which are consistent with our numerical results. Based on our findings and Micnas' work \cite{Micnas1988PRB}, the $d$-wave symmetry arises from the competition among various non-local pairing channels, with their instability driven by the nearest-neighbor attractions. In the context of spin-fluctuation theory \cite{Bickers1989}, the SC instability arising from strong local repulsion is also characterized by a $d$-wave symmetry in the order parameter. 

\textit{Proportionality $T_{c}\propto\left|V\right|$}. Although not observed in cuprates, evidence of such a proportional relationship exists at the $\text{FeSe}/\text{SrTiO}_{3}$ surface \cite{Song2019}. The BCS theory on a flat-band lattice is theoretically expected to result in this proportionality \cite{Miyahara2007}, and our work provides further support for this hypothesis. We believe that two key factors contributing to high $T_c$ are the presence of a broad flat band near the Fermi surface and a sufficiently strong effective intersite attraction.

\textit{Existence of flat band}. The hole-doped cuprate superconductors exhibit partial flat bands near Fermi surface \cite{Shen1995}, possibly originating from vHS. Our numerical results also demonstrate the presence of flat bands and underscore their fundamental role in enhancing $T_c$. Moreover, many unconventional superconductors exhibit flat band characteristics \cite{Cao2018,Shaginyan2022}, implying that flat bands may be a common feature in unconventional superconductors. 

\textit{Coexistence of AFM and SC orders}. Since we assumed an AFM tendency in constructing the mean-field equations, the SC order emerges in the presence of AFM order within a broad range of doping levels. The coexistence of these two orders was also early investigated in Ref. \cite{Tobijaszewska2005} at the mean-field level. This result differs from the phase diagram for hole-doped cuprates \cite{Sobota2021RMP}, where AFM order only exists at lightly doping levels (less than $0.05$, usually). Although, strong spin fluctuations are believed to exist within a broad range of doping levels.

\textit{Towards pairing density wave}. The AFM order breaks the spin $SU\left(2\right)$ and spatial symmetries, resulting the SC order subject to spatial modulation, i.e., $\Delta_{\boldsymbol{\delta},\boldsymbol{r}}^{\prime}=\Delta_{\boldsymbol{\delta},\boldsymbol{0}}^{\prime}+\Delta_{\boldsymbol{\delta},\boldsymbol{Q}}^{\prime}\text{e}^{\text{i}\boldsymbol{Q}\cdot\boldsymbol{r}}$. The nonzero value of $\Delta_{\boldsymbol{\delta},\boldsymbol{Q}}^{\prime}$ corresponds to the pairing operator $\hat{c}_{\downarrow,\boldsymbol{k}} \hat{c}_{\uparrow,\boldsymbol{Q}-\boldsymbol{k}}$, indicating the pairing density wave with momentum $\boldsymbol{Q}$ \footnote{In Ref. \cite{Tobijaszewska2005}, $\Delta_{\boldsymbol{\delta},\boldsymbol{Q}}^{\prime}$ was interpreted as the order parameter for the $\pi$-triplet. However, since the $SU\left(2\right)$ symmetry is broken, $\Delta_{\boldsymbol{\delta}, \boldsymbol{Q}}^{\prime}$ should not refer to a triplet}. In our mean-field calculations, $\Delta^{\prime}_{\boldsymbol{Q}}$ is nonzero only when $m \ne 0$, implying that the pairing density wave only emerges in the presence of the AFM order. We may speculate that in a more precise computation of the extended Hubbard model, pairing density waves arise from strong spin-density-wave fluctuations.

\textit{Towards pseudogap}. When both AFM (antiferromagnetic) and SC (superconducting) orders coexist, two distinct single-particle gaps emerge: the SC gap and the AFM gap. The pseudogap is commonly considered a precursor to an actual gap, and can be generated by both SC and spin fluctuations \cite{Chen2005,Schafer2021}. In the extended Hubbard model, it is possible for two types of pseudogaps to coexist.

\section{Summary}\label{sec:summary}
In summary, we have explored the SC features on the extended Hubbard model at the mean-field level, assuming the system's inclination towards AFM order. Our numerical investigations have revealed several key features, including a predisposition towards $d$-wave pairing, the presence of a dome-shaped dependence of $T_c$ on doping, an upper limit of $0.25\left|V\right|$ for $T_c$ with increasing $U$, and a nearly proportional relationship between $T_c$ and $\left|V\right|$. By examining the system's density of states, we identified the existence of vHS near the Fermi surface, which may serve as a prominent source of high-temperature superconductivity. These results suggest that the extended Hubbard model could be the appropriate framework for investigating cuprate superconductivity, and are believed to offer insights for more precise calculations within this model in future.

Based on our mean-field results, the presence of flat bands and strong electron-phonon interactions may be two crucial factors contributing to the high $T_c$. However, if so, there are several issues that warrant further investigation. These include completing a comprehensive mean-field analysis \cite{Kato1990}, considering higher-order perturbation theories \cite{Bickers1989}, and incorporating dynamic effective attractions \cite{Wang2021PRL}. It is crucial to note that flat bands combined with strong interactions can introduce various correlated effects, and simple perturbation theory may not yield quantitatively satisfactory results. Further research is needed to address these challenges adequately.

\bibliography{main.bib}
\onecolumngrid
\appendix
\section*{Gap equation for pure $s$ wave pairing}

For pure $s$-wave pairing, the system of gap equations is given by
\begin{subequations}
\begin{align}
X_{\boldsymbol{0},\gamma} & =+\frac{\left|V\right|}{\mathcal{V}}\sum_{\boldsymbol{p}^{\prime}}\gamma_{\boldsymbol{p}^{\prime}}^{2}\left(F_{\boldsymbol{p}^{\prime}}^{-}+F_{\boldsymbol{p}^{\prime}}^{+}\right)X_{\boldsymbol{0},\gamma} \nonumber \\
& \quad -\frac{\left|V\right|}{\mathcal{V}}\sum_{\boldsymbol{p}^{\prime}}\gamma_{\boldsymbol{p}^{\prime}}^{2}\left(F_{\boldsymbol{p}^{\prime}}^{-}-F_{\boldsymbol{p}^{\prime}}^{+}\right)\sin\theta_{\boldsymbol{p}^{\prime}}X_{\boldsymbol{Q},\gamma}\nonumber \\
 & \quad-\frac{U}{\mathcal{V}}\sum_{\boldsymbol{p}^{\prime}}\gamma_{\boldsymbol{p}^{\prime}}\left(F_{\boldsymbol{p}^{\prime}}^{-}-F_{\boldsymbol{p}^{\prime}}^{+}\right)\cos\theta_{\boldsymbol{p}^{\prime}}\Delta, \\
X_{\boldsymbol{Q},\gamma} & =-\frac{\left|V\right|}{\mathcal{V}}\sum_{\boldsymbol{p}^{\prime}}\gamma_{\boldsymbol{p}^{\prime}}^{2}\sin\theta_{\boldsymbol{p}^{\prime}}\left(F_{\boldsymbol{p}^{\prime}}^{-}-F_{\boldsymbol{p}^{\prime}}^{+}\right)X_{\boldsymbol{0},\gamma}\nonumber \\
& \quad +\frac{\left|V\right|}{\mathcal{V}}\sum_{\boldsymbol{p}^{\prime}}\gamma_{\boldsymbol{p}^{\prime}}^{2}\left(\sin^{2}\theta_{\boldsymbol{p}^{\prime}}\left(F_{\boldsymbol{p}^{\prime}}^{-}+F_{\boldsymbol{p}^{\prime}}^{+}\right)+2\cos^{2}\theta_{\boldsymbol{p}^{\prime}}F_{\boldsymbol{p}^{\prime}}^{\prime}\right)X_{\boldsymbol{Q},\gamma}\nonumber \\
 & \quad+\frac{U}{\mathcal{V}}\sum_{\boldsymbol{p}^{\prime}}\gamma_{\boldsymbol{p}^{\prime}}\left(F_{\boldsymbol{p}^{\prime}}^{-}+F_{\boldsymbol{p}^{\prime}}^{+}-2F_{\boldsymbol{p}^{\prime}}^{\prime}\right)\sin\theta_{\boldsymbol{p}^{\prime}}\cos\theta_{\boldsymbol{p}^{\prime}}\Delta, \\
\Delta & =+\frac{\left|V\right|}{\mathcal{V}}\sum_{\boldsymbol{p}^{\prime}}\left(F_{\boldsymbol{p}^{\prime}}^{-}-F_{\boldsymbol{p}^{\prime}}^{+}\right)\cos\theta_{\boldsymbol{p}^{\prime}}X_{\boldsymbol{0},\gamma}\gamma_{\boldsymbol{p}^{\prime}}\nonumber \\
& \quad-\frac{\left|V\right|}{\mathcal{V}}\sum_{\boldsymbol{p}^{\prime}}\left(F_{\boldsymbol{p}^{\prime}}^{-}+F_{\boldsymbol{p}^{\prime}}^{+}-2F_{\boldsymbol{p}^{\prime}}^{\prime}\right)\sin\theta_{\boldsymbol{p}^{\prime}}\cos\theta_{\boldsymbol{p}^{\prime}}X_{\boldsymbol{Q},\gamma}\gamma_{\boldsymbol{p}^{\prime}}\nonumber \\
 & \quad-\frac{U}{\mathcal{V}}\sum_{\boldsymbol{p}^{\prime}}\left(\cos^{2}\theta_{\boldsymbol{p}^{\prime}}\left(F_{\boldsymbol{p}^{\prime}}^{-}+F_{\boldsymbol{p}^{\prime}}^{+}\right)+2\sin^{2}\theta_{\boldsymbol{p}^{\prime}}F_{\boldsymbol{p}^{\prime}}^{\prime}\right)\Delta.\label{eq:s-wave-eqs}
\end{align}
\end{subequations}
For most parameters in consideration, $T_c$ for $s$ wave pairing instability is smaller than those for $d$ and $p$ pairing instabilities. This is because the $s$ wave pairing instability will be suppressed by the local repulsion. 
\end{document}